\documentclass[10pt,twocolumn,superscriptaddress,english,prb,aps,longbibliography]{revtex4-2}
\usepackage[utf8]{inputenc}
\usepackage{amsmath}
\usepackage{amssymb}
\usepackage{bm}
\usepackage{graphicx}
\usepackage{xspace}
\usepackage{siunitx}
\usepackage{mathtools}
\usepackage{xcolor}
\usepackage{bbold}
\makeatletter
\@ifundefined{textcolor}{}
{\definecolor{BLACK}{gray}{0}
 \definecolor{WHITE}{gray}{1}
 \definecolor{RED}{rgb}{1,0,0}
 \definecolor{GREEN}{rgb}{0,1,0}
 \definecolor{BLUE}{rgb}{0,0,1}
 \definecolor{CYAN}{cmyk}{1,0,0,0}
 \definecolor{MAGENTA}{cmyk}{0,1,0,0}
 \definecolor{YELLOW}{cmyk}{0,0,1,0}
}

\usepackage[normalem]{ulem}

\usepackage{soul}
\usepackage{braket}
\usepackage[backref=none,
bookmarksnumbered=true,
bookmarks=true,
bookmarksopen=true,
colorlinks=true,
citecolor=blue,
linkcolor=blue,
anchorcolor=green,
urlcolor=blue,unicode=false]{hyperref}

\newcommand{\add}[1]{#1} 
 
\def\localpha{m_1}
\def\locbeta{m_2}
\newcommand\phitb{\phi^{TB}}
\newcommand{\km}{|\mathbf K_M|}
\newcommand{\khopping}{t}
\newcommand{\rydberg}{E_0}
\newcommand{\mote}{{MoTe\textsubscript{2}\,}}
\newcommand{\wse}{{WSe\textsubscript{2}}}

\newcommand{\auc}{A_{\text{UC}}}
\newcommand{\timereversal}{T}

\newcommand{\beff}{\tilde B} 
\newcommand{\hc}{\text{h.c.}}

\def\localpha{m_1}
\def\locbeta{m_2}

\begin{document}
\begin{center}

\title{Hofstadter spectrum of 
Chern bands in twisted transition
metal dichalcogenides}

\author{Kry\v{s}tof Kol\'{a}\v{r}}
\affiliation{\mbox{Dahlem Center for Complex Quantum Systems and Fachbereich Physik, Freie Universit\"at Berlin, 14195 Berlin, Germany}}
\author{Kang Yang}
\affiliation{\mbox{Dahlem Center for Complex Quantum Systems and Fachbereich Physik, Freie Universit\"at Berlin, 14195 Berlin, Germany}}
\author{Felix von Oppen}
\affiliation{\mbox{Dahlem Center for Complex Quantum Systems and Fachbereich Physik, Freie Universit\"at Berlin, 14195 Berlin, Germany}}
\author{Christophe Mora}
\affiliation{Universit\'e Paris Cit\'e, CNRS,  Laboratoire  Mat\'eriaux  et  Ph\'enom\`enes  Quantiques, 75013  Paris,  France}

\date{\today}
\begin{abstract}
We study the topological bands in twisted 
bilayer transition metal dichalcogenides in an external magnetic field.
We first focus on a paradigmatic model of \wse{}, 
which can be described in an adiabatic approximation 
as particles moving in a periodic potential and an emergent periodic magnetic field with nonzero average.
We understand the magnetic-field dependent spectra of \wse{} based on the point \textit{net zero flux}, at which the external field  cancels the average emergent field. 
At this point, the band structure interpolates between the tightly-bound and nearly-free (weak periodic potential) paradigms as the twist angle increases. For small twist angles, the energy levels in a magnetic field mirror 
the Hofstadter butterfly of the Haldane model. For larger twist angles, the isolated Chern band at zero flux
evolves from nearly-free bands at the point of \textit{net zero flux}.
We also apply our framework to a realistic model of twisted bilayer \mote{}, which has recently been suggested to feature higher Landau level analogs. We show that at negative unit flux per unit cell, the bands exhibit remarkable similarity to a backfolded parabolic dispersion, even though the adiabatic approximation is inapplicable. 
This backfolded parabolic dispersion naturally explains the similarity of the Chern bands at zero applied flux to the two lowest Landau levels, offering a simple picture supporting the emergence of non-Abelian states in twisted bilayer \mote{}.
We propose the study of magnetic field dependent band structures as a versatile method to investigate the nature of topological bands 
and identify Landau level analogs. 

\end{abstract}
\maketitle 

\end{center}

\section{Introduction}
When magnetic flux of the order of a flux quantum $\Phi_0=h/e$ threads the unit cell of a lattice, the energy spectrum 
exhibits a remarkable fractal structure, first studied by Hofstadter \cite{hofstadterHofstadterEnergyLevelsWave1976}.
Reaching this regime for standard materials requires unrealistically large applied magnetic fields $B$. 
The advent of moir\'e materials which feature large unit cells
enabled the Hofstadter regime to be probed experimentally,
with signatures observed in bilayer graphene on hBN \cite{kimDeanHofstadterButterflyFractal2013},
twisted bilayer graphene \cite{feldmanYuCorrelatedHofstadterSpectrum2022,youngSaitoHofstadterSubbandFerromagnetism2021,efetovDasObservationReentrantCorrelated2022}, and transition metal dichalcogenide heterobilayers \cite{feldmanKometterHofstadterStatesReentrant2023}.

The experimental realization of twisted bilayer graphene \cite{jarillo-herreroCaoCorrelatedInsulatorBehaviour2018,jarillo-herreroCaoCorrelatedInsulatorBehaviour2018} 
spearheaded these developments \cite{deanYankowitzTuningSuperconductivityTwisted2019,kimHaoElectricFieldTunable2021,yazdaniOhEvidenceUnconventionalSuperconductivity2021,efetovLuSuperconductorsOrbitalMagnets2019,jarillo-herreroCaoNematicityCompetingOrders2021,liLiuTuningElectronCorrelation2021,nadj-pergeAroraSuperconductivityMetallicTwisted2020,efetovStepanovUntyingInsulatingSuperconducting2020,youngSaitoIndependentSuperconductorsCorrelated2020,ilaniZondinerCascadePhaseTransitions2020a,yazdaniWongCascadeElectronicTransitions2020}. Twisted bilayer graphene 
exhibits flat bands at twist angles close to the magic angle of $\theta = 1.05^\circ$ \cite{macdonaldBistritzerMoireBandsTwisted2011}, providing additional interest in its Hofstadter physics. The band structure at large magnetic fields was shown to exhibit rich structure \cite{macdonaldBistritzerMoireButterfliesTwisted2011,balentsHejaziLandauLevelsTwisted2019,bernevigHerzog-ArbeitmanReentrantCorrelatedInsulators2022,sternShefferChiralMagicangleTwisted2021},
with salient features appearing at magnetic fields corresponding to an integer number of flux quanta threading the unit cell.
At these points, the Hofstadter problem retains the periodicity of the underlying lattice \cite{bernevigHerzog-ArbeitmanReentrantCorrelatedInsulators2022},
and the system displays reentrant flat bands, resulting in an interaction-driven phenomenology, which is similar to that at zero magnetic field \cite{efetovDasObservationReentrantCorrelated2022}.
Shortly after twisted bilayer graphene, twisted transition metal dichalcogenide (TMD) bilayers emerged as another remarkable moir\'e platform,
featuring flat topological bands \cite{macdonaldWuTopologicalInsulatorsTwisted2019,fuDevakulMagicTwistedTransition2021}
\add{\cite{queirozCrepelChiralLimitOrigin2024,franzZhouMoirFlatChern2022}}
and permitting the observation of Mott insulators \cite{deanWangCorrelatedElectronicPhases2020}, 
superconductivity \cite{makXiaUnconventionalSuperconductivityTwisted2024,deanGuoSuperconductivityTwistedBilayer2024},
quantum criticality \cite{pasupathyGhiottoQuantumCriticalityTwisted2021}, and, most remarkably, the integer and fractional quantum anomalous Hall effects 
\cite{xuCaiSignaturesFractionalQuantum2023,xuParkObservationFractionallyQuantized2023,shanZengThermodynamicEvidenceFractional2023,liXuObservationIntegerFractional2023,feldmanFouttyMappingTwisttunedMultiband2024}. Very recently, a TMD bilayer has been proposed to exhibit a fractional quantum spin Hall effect of holes 
\cite{makKangEvidenceFractionalQuantum2024} and shown to feature multiple flat bands of equal Chern number in a given valley
\cite{makKangEvidenceFractionalQuantum2024,makKangObservationDoubleQuantum2024}. 
These experimental findings were accompanied by intense theoretical efforts. Due to flavor polarization, topological bands exhibit an anomalous Hall effect at integer fillings
\cite{wuJiaMoirFractionalChern2024,zaletelWangTopologyMagnetismCharge2023,regnaultYuFractionalChernInsulators2024,wuLiElectricallyTunedTopology2024}, 
with magnetic field tuning the fine balance between competing states \cite{vafekWangInteractingPhaseDiagram2024}. The fractional quantum anomalous
Hall effect can form in partially filled Chern bands and is now firmly established in exact diagonalization studies 
\cite{linLiSpontaneousFractionalChern2021,fuCrepelAnomalousHallMetal2023,regnaultYuFractionalChernInsulators2024,soluyanovAndrewsFractionalQuantumHall2020,fuReddyFractionalQuantumAnomalous2023,xiaoWangFractionalChernInsulator2024,zhangMaoTransferLearningRelaxation2024}, greatly broadening the scope of fractional states in twisted graphenes \cite{bergholtzAbouelkomsanParticleHoleDualityEmergent2020}. 

Motivated by these exciting developments, we study the effects of a strong
magnetic field on topological bands in twisted TMDs. We employ a continuum model~\cite{macdonaldWuTopologicalInsulatorsTwisted2019,fuDevakulMagicTwistedTransition2021} for each valley of the twisted-TMD band structure and develop a gauge-independent framework to compute the spectrum as a function of magnetic field (Hofstadter butterfly). While our framework is generally applicable, we focus specifically on models of twisted bilayer \wse{} and \mote. 

We first investigate a minimal model of \wse{}, identifying and elucidating a remarkable similarity between the Hofstadter butterflies for \wse{} at small twist angles 
and the Haldane model~\cite{haldaneHaldaneModelQuantumHall1988,yangHouNextnearestneighbortunnelinginducedSymmetryBreaking2009}, and shedding light on the various topological phase transitions as a function of magnetic flux. We find that in many ways, the model with an external magnetic field corresponding to $\Phi = -1$ flux quanta per moiré unit cell can be viewed as a natural parent model for understanding the TMD band structures. In particular, we find that up to a twist angle of $\sim 2^\circ$, 
the band structure at $\Phi = -1$ is robustly described by a Haldane model. This  contrasts with the zero-flux case, where a topological phase transition~\cite{fuDevakulMagicTwistedTransition2021} already occurs at a twist angle of $\sim 1.5^\circ$. 
Beyond this transition, the two topmost bands acquire equal Chern numbers, which is no longer compatible with the two-band approximation of the Haldane model. We show that these equal-Chern bands arise from the parabolic band top at flux $\Phi = -1$. 
Notably, the entire topmost band at $\Phi=-1$ resembles a free-particle dispersion, allowing a single Landau level to develop 
into a detached flat Chern band at zero flux, and providing ideal conditions for fractional quantum Hall liquids to form. 

We then apply our framework to a more elaborate model, which provides a realistic description of the band structure of  twisted bilayer \mote{} at twist angle $\theta=2.1^\circ$, close to the value in a recent experiment \cite{makKangEvidenceFractionalQuantum2024}. This model
features three low-lying bands of equal Chern number per valley, in agreement with  experiment. It has been proposed to feature non-Abelian topological order in the second topmost band, analogous to 
the first Landau level \cite{makKangEvidenceFractionalQuantum2024,choAhnFirstLandauLevel2024,xiaoWangHigherLandauLevelAnalogues2024,fuReddyNonAbelianFractionalizationTopological2024,zhangXuMultipleChernBands2024,xiaoZhangPolarizationdrivenBandTopology2024}.
We show that at $\Phi=-1$, the band structure is remarkably close to that of a backfolded free-electron dispersion for a large range of energies, allowing 
its two lowest Landau levels to persist all the way to $\Phi=0$. In this picture, the lowest two bands at $\Phi=0$ are Landau levels of the nearly free electrons at $\Phi=-1$, so that the $\Phi=-1$ band structure may serve as a natural parent model to understand the appearance of non-Abelian phases \cite{makKangEvidenceFractionalQuantum2024,choAhnFirstLandauLevel2024,xiaoWangHigherLandauLevelAnalogues2024,fuReddyNonAbelianFractionalizationTopological2024,zhangXuMultipleChernBands2024,xiaoZhangPolarizationdrivenBandTopology2024}. 

Beyond twisted bilayer TMDs, our results for twisted bilayer \mote{} establish the Hofstadter spectra as a valuable general characterization method for Chern bands and their connection to Landau levels,
complementing earlier approaches focusing on their quantum geometry \cite{royRoyBandGeometryFractional2014,yangWangExactLandauLevel2021,vishwanathLedwithFractionalChernInsulator2020,bergholtzAbouelkomsanQuantumMetricInduced2023,ledwithFujimotoHigherVortexabilityZero2024,wangLiuTheoryGeneralizedLandau2024,liuWangOriginModelFractional2023,crepelEstienneIdealChernBands2023}

\begin{figure}[t]
    \centering
    \includegraphics[width=1\columnwidth]{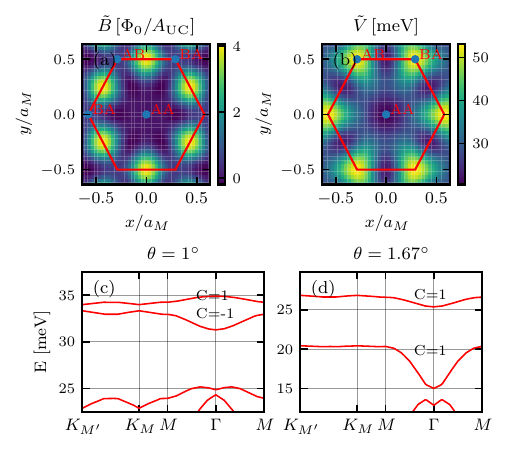}
    \caption{
(a)  Effective magnetic field $\beff(\mathbf r)  =\nabla \times \tilde A(\mathbf r)$ entering the adiabatic model in Eq.~\eqref{eq:sphammacdo}), measured 
in units of flux quanta ($\Phi_0=h/e$) per unit cell ($A_\text{UC}$). The moir\'e unit cell with the AA, AB, and BA stacking locations is highlighted in red. The effective magnetic field is peaked at the midpoints between AB and BA stacking.
(b) Effective moir\'e potential $\tilde V(\mathbf r)$ entering the adiabatic model. (c) Band structure of twisted bilayer \wse{}. Plots in panels (a-c) are for a twist angle of $\theta =1^\circ$ using model parameters from Ref.~\cite{fuDevakulMagicTwistedTransition2021}.
(d) Same as (c) but at a twist angle of $\theta =1.67^\circ$. }
    \label{fig:figone}
\end{figure}

The special role played by the $\Phi=-1$ band structure is best motivated by 
the adiabatic picture of twisted bilayer TMDs 
\cite{yaoZhaiTheoryTunableFlux2020,macdonaldMorales-DuranMagicAnglesFractional2024}. This model describes the band structure of one valley using a model of electrons with effective mass $m^*$ subject to a potential $\tilde V(\mathbf r)$ and an effective magnetic field $\beff(\mathbf
r) = \nabla \times \tilde {\boldsymbol{A}}(\mathbf r)$, both of which are moir\'e periodic. The corresponding Hamiltonian for the $K$ valley takes the form 
\begin{equation}
\label{eq:sphammacdo}
H^{K}_{\text{Adiabatic}}= -\frac{(\hbar \mathbf k-e\tilde{\mathbf{A}}(\mathbf r))^2}{2m^*} + \tilde V(\mathbf r).
\end{equation}
The valley-odd effective magnetic field emerges after projecting to the low-energy states of the potential and tunneling terms of the full continuum Hamiltonian,
exploiting their slow spatial variation. These can be written as
an effective Zeeman field acting on the layer degree of freedom, with the direction  $\mathbf{\hat n}(\mathbf{r})$
describing a moir\'e periodic texture. The effective magnetic field, shown in Fig.~\ref{fig:figone}a, is given by the Pontryagin-index density associated with this skyrmion-like texture, \begin{equation}\label{eq:beffdefmain}
   \beff(\mathbf r) =  -\frac{\hbar}{2e} \mathbf{\hat n}(\mathbf{r})\cdot \partial_x \mathbf{\hat n}(\mathbf{r})\times \partial_y \mathbf{\hat n}(\mathbf{r}),
\end{equation}
and corresponds to an average of one flux quantum per unit cell. 
The moir\'e-periodic scalar potential $\tilde V(\mathbf r)$ experienced by the electrons, shown in Fig.~\ref{fig:figone}b, combines the effective Zeeman energy and a scalar potential originating from the adiabatic approximation, 
\begin{equation}
D(\mathbf{r}) = \frac{\hbar^2}{8m^*}\sum_{i=x,y}[\partial_{i} \mathbf{\hat n}(\mathbf{r})]^2,
\end{equation}
which is proportional to the trace of the quantum geometric tensor of the adiabatically polarized states as a function of $\mathbf{r}$. 

Within the adiabatic model, the external magnetic field simply adds to the effective field, $\tilde B(\mathbf r) \to  B_{\text{tot}}(\mathbf r) = \tilde B(\mathbf r) + B $
\cite{macdonaldMorales-DuranMagicAnglesFractional2024}. This immediately
implies that on average, an externally applied field corresponding to $\Phi=-1$ flux quanta per unit cell cancels the effective magnetic field. At this \textit{point of net zero flux}, the model \add{in a single valley} reduces to a band structure without average magnetic field and can be solved using a Bloch basis for the original lattice unit cell.
This simplification makes $\Phi=-1$ a natural starting point to understand the magnetic-field-dependent spectrum. 
\add{Note that in the other valley, the effective magnetic field $\tilde B(\mathbf r)$ carries the opposite sign, so a cancellation in that valley occurs for opposite applied flux $\Phi=1$ .} 

Within the adiabatic picture, it is natural to describe the band structure of twisted bilayer \wse{} at $\Phi=-1$ in terms of a Haldane model. The periodic potential $\tilde V(\mathbf r)$ localizes  particles into site orbitals on a hexagonal lattice, cf.\ Fig.~\ref{fig:figone}b. The next-nearest neighbor hopping between these orbitals will have nonzero phases due to the inhomogeneous part of the emergent magnetic field.
The departure of the external magnetic flux from  $\Phi=-1$, i.e., the {\it net flux}, tunes these phases. Indeed, for sufficiently low twist angles,
we find very good agreement between the exact magnetic-field spectra obtained
from the original continuum model and those computed from the Haldane model.
Specifically, both models exhibit a similar sequence of topological phase
transitions, which are expected whenever the phase of the next-nearest neighbor
hopping amplitude changes sign.
The site orbitals delocalize with increasing twist angle. As a result, the
Haldane model must be extended to include longer-range hoppings and becomes
more sensitive to the net flux. A direct consequence is that the Haldane model
at {\it net zero flux} ($\Phi=-1$) is much more robust against an increase in the twist angle than at
$\Phi=0$. 

For larger twist angles, the adiabatic model suggests an alternative description at $\Phi=-1$.
The kinetic term in Eq.~\eqref{eq:sphammacdo} becomes dominant, corresponding to holes in a weak periodic potential. The resulting band structure naturally resembles backfolded free particle bands with small gaps. Even though the adiabatic picture breaks down upon the increasing twist angle~\cite{yaoZhaiTheoryTunableFlux2020,macdonaldMorales-DuranMagicAnglesFractional2024,macdonaldShiAdiabaticApproximationAharonovCasher2024}, we still find good agreement with a parabolic band for the topmost band of \wse{} at $\theta=1.67^\circ$.
For \mote{}, fast oscillations in the effective Zeeman field direction $\mathbf{\hat n}(\mathbf{r})$ render
the adiabatic approximation inapplicable, yet the two topmost bands at the point of {\it net zero flux} ($\Phi=-1$) are nearly parabolic.

Our manuscript is structured as follows. In Sec.~\ref{sec:bandstructuresintro}, we introduce the continuum TMD model and outline its solution in the presence of an external magnetic field.
In Sec.~\ref{sec:tightbindingregime}, we analyze the model for \wse{} in the small twist angle regime, which is well approximated within a tight-binding approach. Larger twist angles are considered in Sec.~\ref{sec:nearlyfreeregime}. Section~\ref{sec:mote}  investigates a corresponding model for \mote{}, featuring Chern bands that are akin to the first Landau level.
We conclude with a discussion, highlighting the general applicability and experimental relevance of our findings.

\section{Band structures in a magnetic field}
\label{sec:bandstructuresintro}
\subsection{TMD continuum model}
The band structure of twisted homobilayer TMDs in valley $K$ can be described by the continuum Hamiltonian
\cite{macdonaldWuTopologicalInsulatorsTwisted2019,fuDevakulMagicTwistedTransition2021}
\begin{eqnarray}
\label{eq:sphamtmd}
H^{K}_{\text{sp}}&=& \begin{pmatrix}-\frac{\hbar^2 (\mathbf k-\mathbf K^b)^2}{2m^*} + V_b(\mathbf r) & T(\mathbf r)  \\ 
T^\dagger(\mathbf r)  &-\frac{\hbar^2 (\mathbf k-\mathbf K^t)^2}{2m^*} + V_t(\mathbf r)
\end{pmatrix},\quad
\end{eqnarray}
where $m^*$ is the effective mass of the valence band,
$\mathbf K^{t/b} = (0,\pm 4\pi\sin(\theta/2)/(3a_0 ))$
are the momenta of the band extrema of the top and bottom layers with $a_0$ the TMD lattice constant,
$V_{t/b}(\mathbf r)$ are the potentials in the top/bottom layers,
and $T(\mathbf r)$ describes interlayer tunneling. In the first harmonic approximation, these take the form
\begin{eqnarray}
V_b(\mathbf r) &=& Ve^{i\psi} (e^{i\mathbf g_1 \cdot \mathbf r}+e^{i\mathbf g_3 \cdot \mathbf r}+e^{i\mathbf g_5 \cdot \mathbf r})   +\hc \\
\label{eq:vtvbrelation}
V_t(\mathbf r) &=& V_b(-\mathbf r) \\
T(\mathbf r)& =& w (1 + e^{-i \mathbf g_2 \cdot \mathbf r} + e^{-i \mathbf g_3 \cdot \mathbf r}),
\end{eqnarray}
with tunneling strength $w$, potential strength $V$, the phase $\psi$, and the six reciprocal vectors $\mathbf g_j$ defined as the $j-1$ counterclockwise $C_{6z}$ rotations of $\mathbf g_1 = (4\pi \theta /(\sqrt{3}a_0),0)$. 
The moir\'e lattice constant $a_M = \frac{a_0}{2\sin(\theta/2)}$ is much larger than the bare TMD lattice constant $a_0$. 

The form of the Hamiltonian is fixed by the $C_{3z}$, $C_{2y}$, and time-reversal ($\timereversal$) symmetries of a twisted TMD homobilayer.
In the first harmonic approximation, there is an additonal 3D inversion symmetry acting as $\sigma_x H^{K}_{\text{sp}} (\mathbf r) \sigma_x = H^{K}_{\text{sp}} (-\mathbf r) $, which, however, is broken by higher harmonic terms
\footnote{$C_{2y}$ fixes $V_b(x,y) = V_t(-x,y)$ only. If only first harmonics are considered, this turns out to be the same as $V(\mathbf r) =V(- \mathbf r)$, see Eq.~\eqref{eq:vtvbrelation} in the text.}~\cite{regnaultYuFractionalChernInsulators2024,canoWangChiralApproximationTwisted2021}.
Note that $C_{2y}\timereversal$ remains a good symmetry in the presence of an externally applied magnetic field, even though $C_{2y}$ and $\timereversal$ are both broken individually. 

Many different parameter choices have been proposed in the literature for relevant twisted TMDs
\cite{macdonaldWuTopologicalInsulatorsTwisted2019,fuDevakulMagicTwistedTransition2021,xiaoWangFractionalChernInsulator2024}.
For most of this work, we consider a minimal model of twisted \wse{} in the first harmonic approximation from
Ref.~\cite{fuDevakulMagicTwistedTransition2021} as an illustrative example.
The parameters of this model are $(a_0,m^*,V,\psi,w) = (\SI{0.332}{nm} ,\SI{0.43 }{m_e},\SI{9}{meV},-128^\circ, \SI{18}{meV}) $,
with $m_e$ the bare electron mass.
The band structure of twisted \wse{} depends sensitively on twist
angle. At small twist angles, 
it has two low-energy
Chern bands of opposite Chern number separated from other bands by a large gap, as shown for $\theta\approx 1^\circ$ in Fig.~\ref{fig:figone}c.
In this regime, the
two bands are accurately described by a Haldane model \cite{macdonaldWuTopologicalInsulatorsTwisted2019,millisCrepelBridgingSmallLarge2024}.
As the twist angle is increased, the lower of the two bands approaches the remote bands, while the upper band flattens, reaching the magic angle at $\theta\approx 1.43^\circ$ \cite{fuDevakulMagicTwistedTransition2021}.
After a band crossing at $\theta\approx 1.5^\circ$, the Chern number of the lower band switches sign, so that
the two topmost bands have equal Chern number.
We illustrate the band structure in this regime in Fig.~\ref{fig:figone}d for $\theta =1.67^\circ$.

In Sec.~\ref{sec:mote}, we extend our discussion to a realistic model of \mote{} at $\theta=2.1^\circ$, which includes higher harmonics for the tunneling and layer potential terms \cite{choAhnFirstLandauLevel2024,xiaoZhangPolarizationdrivenBandTopology2024}.
Specifically, we add the second harmonic potential
$V^{(2)}_b(\mathbf r) = V^{(2)}_t(\mathbf r)= V_2\left[e^{i\mathbf (\mathbf g_1+\mathbf g_2) \cdot \mathbf r}+e^{i (\mathbf g_3+\mathbf g_4) \cdot \mathbf r}+e^{i(\mathbf g_5+\mathbf g_6) \cdot \mathbf r}\right]   +\hc$ to $V_b(\mathbf r)$ and $V_t(\mathbf r)$ as well as the second harmonic tunneling
$T^{(2)}(\mathbf r) = w_2 \left[e^{i \mathbf g_1 \cdot \mathbf r} + e^{i \mathbf g_4 \cdot \mathbf r} + e^{i (\mathbf g_3+ \mathbf g_2) \cdot \mathbf r}\right]$ to $T(\mathbf r)$.
We use parameters  $(a_0,m^*,V,\psi,w,V_2,w_2) = (\SI{0.3472}{nm} ,\allowbreak \SI{0.62}{m_e},\SI{20.51}{meV},-61.49^\circ,  \allowbreak \SI{-7.01}{meV},\allowbreak \SI{-9.08}{meV},\allowbreak \SI{11.08}{meV})$ 
\cite{choAhnFirstLandauLevel2024}, which fit  the band structure obtained from ab initio calculations \cite{xiaoZhangPolarizationdrivenBandTopology2024}.

\begin{figure*}[t]
    \centering
    \includegraphics[width=1.9\columnwidth]{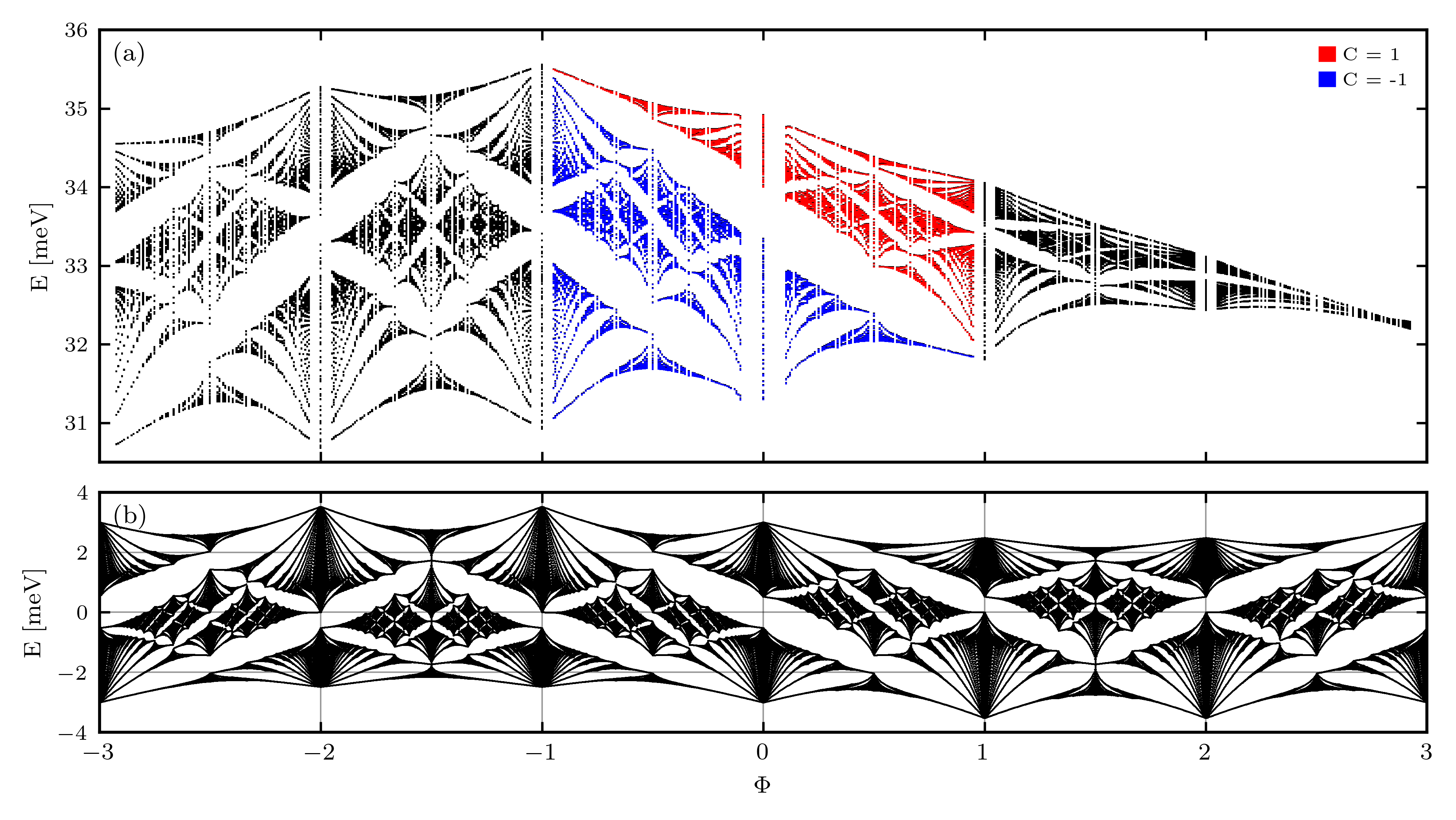}
    \caption{ (a) Energy level diagram for twisted bilayer \wse{} vs.\ applied magnetic field at twist angle $\theta =1^\circ$.
    The topmost $C=1$ band is highlighted in red and the second topmost $C=-1$ band in blue.
(b) Energy level diagram for the topological Haldane model with next-nearest-hopping phase $\phitb= -\pi/6$ at $\Phi=-1$. }
    \label{fig:figbutter}
\end{figure*}

\subsection{Gauge-independent calculation of magnetic Bloch bands}

We introduce an external magnetic field $B$ into the continuum model in Eq.~\eqref{eq:sphamtmd} by 
substituting momentum with kinetic momentum, 
$\mathbf k \to \mathbf \Pi = \mathbf k - e\mathbf{A}/\hbar$. The kinetic momentum obeys
\begin{equation}
\left[\Pi_x,\Pi_y \right]= i eB/\hbar = \frac{i}{l_B^2},
\end{equation}
where $l_B = \sqrt{\hbar/eB}$ is the magnetic length
satisfying $2\pi l_B^2B= \Phi_0$. Working at a rational flux $\Phi = \frac{p}{q}$ per moir\'e unit cell (measured in units of $\Phi_0$), we have $\frac{1}{l_B^2}=2\pi \frac{p}{q} \frac{1}{\auc}$,
where $\auc = |\mathbf a_1 \times \mathbf a_2|$ is the area of the moir\'e unit cell with elementary lattice vectors $\mathbf a_i $. We denote the basis vectors of the corresponding reciprocal lattice as $\mathbf G_1$ and $\mathbf G_2$, which satisfy $\mathbf a_i \cdot \mathbf G_j =2\pi \delta_{ij}$. (In our numerics, we use $\mathbf{G}_1= \mathbf g_5$ and $\mathbf{G}_2=\mathbf g_1$.) We express the Hamiltonian in the Landau level basis \cite{macdonaldBistritzerMoireButterfliesTwisted2011,balentsHejaziLandauLevelsTwisted2019}, working in a gauge-independent formalism. 
We recall the magnetic translation operators
\begin{equation}
T_{\mathbf a} = \exp \left[ i \mathbf a \cdot \left(\mathbf \Pi + \frac{1}{l_B^2}\hat{\bf z} \times \mathbf r \right) \right] 
\end{equation}
($\hat{\bf z}$ is the unit vector in the $z$-direction), which satisfy 
\begin{equation}
\label{eq:magtransalgebra}
T_{\mathbf a}
T_{\mathbf a'} = \exp \left( i\frac{\mathbf a \times \mathbf a' }{l_B^2} \right) 
T_{\mathbf a'}
T_{\mathbf a}.
\end{equation}
where $\mathbf a \times \mathbf a'$ stands for the (oriented) area spanned by the two
in-plane vectors. The operators $T_{\mathbf a}$ and $T_{\mathbf a'}$ commute if this area is threaded by an integer number of flux quanta. We now choose
translations by $q \mathbf a_1$ and $\frac{1}{p} \mathbf a_2$, which enclose precisely one flux quantum, to define our basis. Specifically, we choose a Landau-level basis of simultaneous eigenstates of 
$\mathbf \Pi^2$, $T_{q \mathbf a_1}$, and $T_{ \mathbf a_2/p}$ \cite{rezayiHaldanePeriodicLaughlinJastrowWave1985}.
The basis states $\ket{n, \mathbf k}$ are characterized by a Landau level index $n$ and momentum $\mathbf k$, defined through 
\begin{eqnarray}
\label{eq:deflatticetranslations}
(\Pi_x^2+\Pi_y^2)  \ket{n,\mathbf k}  =(n+\frac{1}{2}) \frac{1}{l_B^2} \ket{n,\mathbf k} \\
\label{eq:deflatticetranslations1}
T_{q \mathbf a_1} \ket{n,\mathbf k} =  \exp \left(i q \mathbf a_1 \cdot \mathbf k \right)\ket{n,\mathbf k} \\
\label{eq:deflatticetranslations2}
T_{\mathbf a_2/p} \ket{n,\mathbf k} =  \exp \left(i \frac{1}{p}\mathbf a_2 \cdot \mathbf k \right) \ket{n,\mathbf k}.
\end{eqnarray}
We expand the momentum $\mathbf k$ in the basis vectors $\mathbf G_1$ and $\mathbf G_2$ of the reciprocal lattice as 
\begin{equation}
\mathbf k = k_1 \frac{1}{q} \mathbf G_1 + k_2 p \mathbf G_2,
\end{equation}
with coefficients $k_1,k_2 \in [0,1)$ defining the Landau-level Brillouin zone. (Note that, as defined, the Landau-level Brillouin zone is $p$ times larger than the conventional magnetic Brillouin zone.)

We now construct this basis starting from the state invariant under magnetic translations, $\ket{n,0}.$ 
To that end, we use that the exponential of the guiding center operator $\mathbf R = \mathbf r -l^2_B (\hat z \times \mathbf \Pi)$
implements a momentum boost \cite{yangWangExactLandauLevel2021},
and is related to magnetic translations through 
\begin{equation}
\label{eq:relationmagnetictranseiqR}
 T_{l_B^2(\hat{z} \times \mathbf q)} = e^{i \mathbf q \cdot \mathbf R} .
\end{equation} 
We use these operators acting on $\ket{n,0}$ 
to explicitly construct states for any $\mathbf k$ in the Landau-level Brillouin zone,
\begin{equation}
\label{eq:llbasisdef}
    \ket{n,\mathbf k} =  
    e^{i k_2 p \mathbf G_2 \cdot \mathbf R}
    e^{i k_1\frac{1}{q}\mathbf G_1 \cdot \mathbf R}
\ket{n,0},
\end{equation}
where Eqs.~\eqref{eq:deflatticetranslations1} and \eqref{eq:deflatticetranslations2} follow using Eqs.~\eqref{eq:relationmagnetictranseiqR}
and \eqref{eq:magtransalgebra}.
We define states with $\mathbf k$ outside the Landau-level Brillouin zone by the periodic extension $\ket{n,\mathbf k} = \ket{n, \mathbf k+ p \mathbf G_2} = 
\ket{n, \mathbf k+ \frac{1}{q} \mathbf G_1}$. With this convention, the states at the Landau-level Brillouin zone boundaries are not continuous. This reflects the topology of Landau levels as states translated across the entire Landau-level Brillouin zone are related by an irremovable $U(1)$ phase. 

The solution now proceeds by expressing the Hamiltonian in the above basis.
The kinetic term only acts on the Landau level index and is independent of momentum $\mathbf k$.
On the other hand, the moir\'e potential and tunneling terms depend on the momentum,
requiring us to evaluate the matrix element of $\exp(i \mathbf g \cdot \mathbf r)$,
\begin{multline}
\label{eq:matrixelementsep}
\braket{n'\mathbf k'|\exp\left(i \mathbf g \cdot \mathbf r\right)|n \mathbf k} =
\braket{n'|\exp\left(i \frac{\Pi \times \mathbf g}{B}\right)|n } \\
 \braket{\mathbf k'|\exp\left(i \mathbf g \cdot \mathbf R\right)|\mathbf k},
\end{multline}
where we separated the position operator into its kinetic momentum and guiding center components. 
The vector $\mathbf g$ is an element of the moiré reciprocal lattice. The first term in Eq.~\eqref{eq:matrixelementsep} is given in terms of Laguerre polynomials in 
App.~\ref{app:landauformulas}, Eq.~\eqref{eq:laguerreev}. The vector $\mathbf g$ can be expanded using a basis of the reciprocal lattice, $\mathbf g = \localpha  \mathbf G_1 + \locbeta \mathbf G_2$, with integers $\localpha$, $\locbeta$. The second term in Eq.~\eqref{eq:matrixelementsep} is then readily derived using the basis of Eq.~\eqref{eq:llbasisdef},
\begin{multline}
\label{eq:matrixelementfinal}
 \braket{\mathbf k'|\exp\left(i \mathbf g \cdot \mathbf R\right)|\mathbf k} = \delta_{\mathbf k', \mathbf k + m_2 \mathbf G_2} 
\exp \left(- i\pi  m_1 m_2 \frac{q}{p} \right)\\
\exp(-i 2\pi k_1 \lfloor k_2+ m_2/p \rfloor ) 
\exp\left\{i 2\pi  (k_2 p +m_2) \frac{q}{p} m_1\right\},
\end{multline}
as shown in App.~\ref{app:landauformulas}.
This states that the Brillouin zone momentum of the Landau level  is conserved up to multiples of $\mathbf G_2$,
The orbit consists of $p$ momenta $\mathbf k, \mathbf k+ \mathbf G_2, \ldots \mathbf k +(p-1)\mathbf G_2$, which have to be included in 
the calculation.
We emphasize that our approach is completely independent of the electromagnetic 
gauge chosen. Going beyond  Ref.~\cite{bernevigHerzog-ArbeitmanMagneticBlochTheorem2022}, which
also works in a gauge-free manner, we obtain the matrix element [Eq.~\eqref{eq:matrixelementfinal}] in closed form in a much simpler fashion, owing to the advantageous basis construction of Eq.~\eqref{eq:llbasisdef}.
\add{We note that for small fluxes, as well as for fluxes close to simple fractions, the hybrid Wannier function approach of 
\cite{vafekWangNarrowBandsMagnetic2022,vafekWangInteractingPhaseDiagram2024} can be used to improve performance \footnote{Note that Ref.~\cite{vafekWangInteractingPhaseDiagram2024} studies twisted \mote{} at $\theta=3.89^\circ$ under magnetic field to explore effects of interactions. In this case, the single-particle bands at $\Phi=0$ realize a Haldane model. However, the Haldane-like bands are less detached from the remote bands than in \wse{} at $\theta=1^\circ$. Thus, while the band structures are in line with the Haldane model physics exposed in Sec.~\ref{sec:tightbindingregime} for negative flux,  the gap to the remote bands closes at positive flux $\Phi=1$, preventing a Haldane model description. }.}
\section{Twisted \wse{} in a magnetic field: tight binding regime}
\label{sec:tightbindingregime}
\begin{figure*}[t]
    \centering
    \includegraphics[width=1\textwidth]{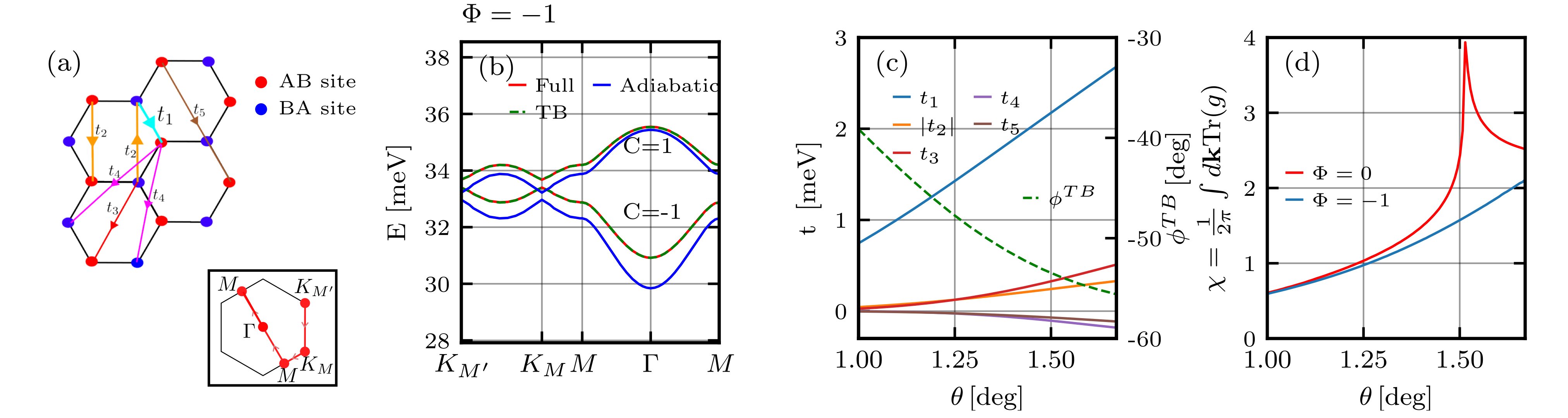}
    \caption{(a) Scheme of the hexagonal lattice model used to fit the two topmost bands of twisted bilayer \wse{} at the \textit{point of net zero flux} ($\Phi=-1$),
showing the hoppings $t_1, t_2,t_3,t_4,t_5$ \add{(top); moir\'e Brillouin zone with the path (red) used to plot band structures (bottom).}
    (b) Two topmost bands of twisted bilayer \wse{} at twist angle $\theta=1^\circ$ and flux $\Phi=-1$. Full band structure is shown in red, adiabatic approximation in blue, and a tight-binding fit 
    to the full band structure in dashed green.
(c) Left axis: Twist angle dependence of the hoppings obtained from the tight-binding fit to the topmost two bands at flux $\Phi=-1$.
Right axis: Phase of next-nearest hopping $\phi^{TB}=\arg(-t_2)$.
(d) $\chi = \frac{1}{2\pi} \int d \mathbf k \mathrm{Tr}(g)$,
the integrated trace of the quantum metric of the topmost two bands at flux $\Phi=-1$ (blue) and $\Phi=0$ (red).}
    \label{fig:figtwo}
\end{figure*}

The first main result of this work, the magnetic field dependent energy spectrum for the continuum model of \wse{} at a twist angle $\theta = 1 ^\circ$, is shown in Fig.~\ref{fig:figbutter}a, focusing on the two topmost detached bands. It has been argued that the spectrum at zero flux is well reproduced by the Haldane model \cite{macdonaldWuTopologicalInsulatorsTwisted2019,fuDevakulMagicTwistedTransition2021},

Figure~\ref{fig:figbutter}a shows that the \textit{point of net zero flux} ($\Phi=-1$) plays a special role. First, the net effective magnetic field vanishes on average at $\Phi=-1$. This is reflected in the fact that the spectrum attains its global maximum at this point. Second, due to the absence of an average magnetic field, the spectrum at this point can be modelled by a Haldane model in its originally envisaged form \cite{haldaneHaldaneModelQuantumHall1988}. The Haldane model
emerges in the adiabatic approximation from orbitals forming a hexagonal lattice localized in the AB and BA regions of the unit cell (Fig.~\ref{fig:figone}b),
with hopping phases arising due to the remnant inhomogeneous part of the effective magnetic field.  
By fitting the two topmost bands with Chern number sequence $(1,-1)$, we obtain a Haldane model with nearest-neighbor hopping $t_1= \SI{0.75}{meV}$ and next-nearest-neighbor hopping $|t_2| = \SI{0.045}{meV}$ with phase $\phitb \approx -40^\circ$, where $t_2 = -e^{i\phitb}|t_2|$. 
Longer range hoppings decay quickly and have little effect on the band structure at this twist angle.

In the Haldane model, Peierls substitution implies that 
the next-nearest-neighbor hopping phase $\phitb$ changes  as $\phitb \to \phitb -  2\pi/6$ as 
an additional flux quantum is threaded, $\Phi \to \Phi+1$. 
Topological phase transitions occur at $\phitb=0,\pi$. 
For our $\Phi=-1$ fit, this line of reasoning would suggest the Chern number sequence to remain $(1,-1)$ for $\Phi=-1,0,1$ and switch to $(-1,1)$ for $\Phi=-3,-2,2$, which agrees with the full continuum-model calculation.
We caution, however, that while the topology agrees with the prediction of the Peierls substitution, the actual best-fit tight binding phases do not follow the predictions of the Peierls substitution, except between $\Phi=-1$ and $\Phi=-2$. This breakdown of the Peierls substitution was highlighted in Ref.~\cite{modugnoIbanez-AzpirozBreakdownPeierlsSubstitution2014} in the context of cold atoms. Nevertheless, we find that the magnetic field dependent energy spectrum of a tight-binding Haldane model \cite{yangHouNextnearestneighbortunnelinginducedSymmetryBreaking2009}, shown in Fig.~\ref{fig:figbutter}b, 
exhibits a striking qualitative resemblance to  Fig.~\ref{fig:figbutter}a.

The behavior between integer fluxes can be understood using the St\v{r}eda formula \cite{stvredaStvredaTheoryQuantisedHall1982}
\begin{equation}
\label{eq:stredaformula}
   \frac{d \nu}{d\Phi} = C,
\end{equation}
which relates Chern number $C$ to the change in the number of states per unit cell $\nu$ inside a band upon varying flux $\Phi$. We illustrate the St\v{r}eda formula by tracking the two bands at zero flux, highlighting the $C=1$ band in red and the second highest $C=-1$ band in blue in Fig.~\ref{fig:figbutter}a.
At $\Phi=1$, the top $C=1$ band expands to $\nu=2$,  whereas the bottom $C=-1$ band goes extinct, necessitating a gap closing since the two bands are detached~\cite{bernevigHerzog-ArbeitmanHofstadterTopologyNoncrystalline2020}. The situation is reversed at $\Phi=-1$, where the $C=-1$ band expands to $\nu=2$, while the $C=1$ band disappears, forcing a gap closing.
The energy level diagram also shows that the Chern bands at zero flux in fact emerge as lowest Landau levels from parabolic band edges, as can be clearly seen near flux $\Phi=1$ for the $C=-1$ band (blue) and near flux $\Phi=-1$ for the $C=1$ band (red). 

Another interesting region is between fluxes $\Phi=-1$ and $\Phi=-2$, since the Chern number sequence is opposite 
at these two fluxes. 
This leads to tension at flux $\Phi=-3/2$ at which the bottom $C=1$ band emerging from $\Phi=-2$ collides with the $C=-1$ band emerging from $\Phi=-1$. More precisely, the St\v{r}eda formula predicts the bottom bands at both $\Phi=-2$ and $\Phi=-1$, with opposite Chern numbers, to expand to $\nu=3/2$ states per unit cell. This necessitates a gap closing. In the continuum model, the gap emerging from $\Phi=-1$ is closing at $\Phi=-3/2$. 
In the Haldane model, we used a symmetric value of $\phitb=-\pi/6$ at $\Phi=-1$, leading to the simultaneous closing of the gaps arising from both
$\Phi=-2$ and $\Phi=-1$, cf.~Fig.~\ref{fig:figbutter}b. 

Remarkably, while at small twist angles, a Haldane description is possible at any applied flux, 
at the \textit{point of net zero flux} ($\Phi=-1$), this description remains valid for an extended range of twist angles.
Longer-range hoppings become necessary to fit the topmost two bands of the adiabatic band structures,
but the character of a hexagonal lattice remains as does the topological character of the Haldane model.
In contrast, at zero flux $\Phi=0$, the emergent magnetic field induces a topological phase transition after which
the bands develop a topology incompatible with a two-band model \cite{millisCrepelBridgingSmallLarge2024}, resulting in two bands of equal Chern number.

To illustrate this point, we fit the band structure at   \textit{net zero flux} ($\Phi=-1$)  
using a hexagonal tight-binding model, illustrated in Fig.~\ref{fig:figtwo}a, retaining up to fifth-nearest-neighbor hoppings.
We present the fit at $\theta = 1^\circ$ in Fig.~\ref{fig:figtwo}b, with the full continuum bands shown in red and the tight-binding fit depicted in dashed green.
We also show the adiabatic band structure in blue, confirming that at this twist angle, the adiabatic approximation is well within its regime of validity. 
In Fig.~\ref{fig:figtwo}c, we plot the hoppings obtained by fitting the bands to the tight-binding model, showing the increase in hopping integrals with twist angle. 
Nevertheless, the model remains relatively short-range. In contrast, at zero flux, $\Phi=0$, the two-band tight binding description becomes invalid
at $\theta \approx 1.5^\circ$ \cite{fuDevakulMagicTwistedTransition2021,millisCrepelBridgingSmallLarge2024}. 
This sharp difference in locality between $\Phi=-1$ and $\Phi=0$ is further brought out 
by looking at the integrated trace of the quantum metric $\chi = \frac{1}{2\pi} \int d \mathbf k \mathrm{Tr}(g)$ of the two topmost bands, shown
in Fig.~\ref{fig:figtwo}d. The average quantum metric measures the localization of wavefunctions \cite{vanderbiltMarzariMaximallyLocalizedGeneralized1997}
and shows that for larger twist angles, the two topmost bands are more localized at \textit{net zero flux} ($\Phi=-1$) than at $\Phi=0$.
In fact, at $\Phi=0$, upon crossing the topological phase transition, $\chi$ diverges logarithmically~\cite{ryuMatsuuraMomentumSpaceMetric2010,chendeSousaMappingQuantumGeometry2023}.

\section{Larger twist angles: towards nearly free electrons}
\label{sec:nearlyfreeregime}

\begin{figure}[t]
    \centering
    \includegraphics[width=1\columnwidth]{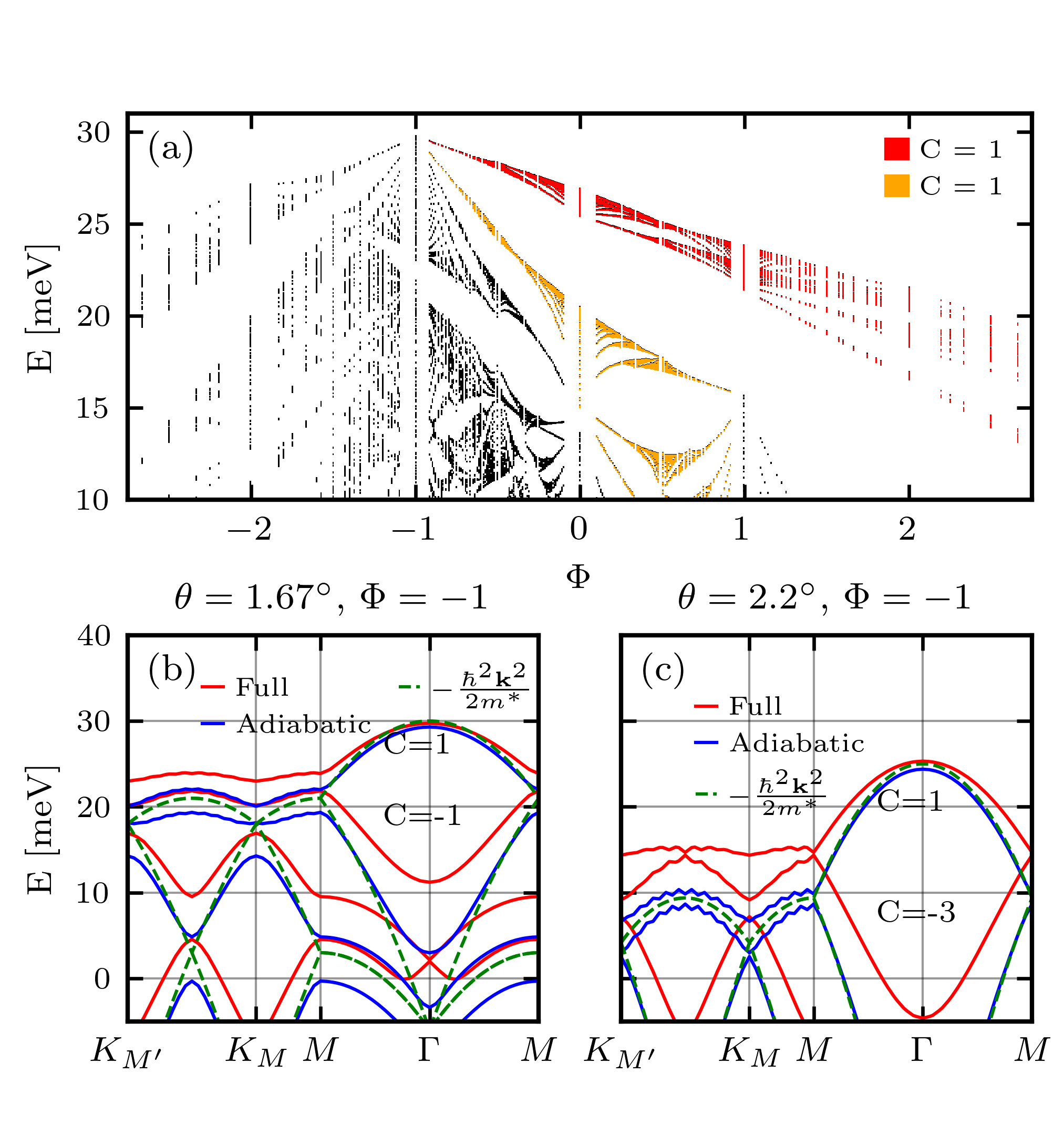}
    \caption{ (a)  Energy level diagram as a function of an applied magnetic field for twisted bilayer \wse{} at twist angle $\theta =1.67^\circ$, showing the topmost $C=1$ band in red and second topmost $C=1$ band in orange.
(b) Band structure at $\theta =1.67^\circ$ and flux $\Phi=-1$. 
The full band structure is shown in red, adiabatic in blue, and a backfolded free electron dispersion in dashed green.
\add{Note that the topmost branch of the backfolded free electron bands are doubly degenerate along the $K_{M'}-K_M-M$ line in the Brillouin zone, cf. the discussion in Appendix~\ref{app:parabolicdegeneracies}.}
(c) Same as (b), but at $\theta =2.2^\circ$. }
\label{fig:figthree}
\end{figure}

The Hofstadter spectrum at a larger twist angle of $\theta = 1.67^\circ$
(see  Fig.~\ref{fig:figthree}a) differs  dramatically from the one at $\theta=1^\circ$ in Fig.\ \ref{fig:figbutter}a. 
The two topmost bands at zero flux now have the same Chern number $C=1$ (cf.\ Fig.~\ref{fig:figone}d), so that the
St\v{r}eda formula, Eq.~\eqref{eq:stredaformula}, predicts their filling $\nu$ to vary in the same way. Both disappear at $\Phi=-1$ and the topmost band (shown in red) expands to two states per unit cell at flux $\Phi=1$ without any
enforced gap closing as in the tight-binding regime discussed above.

We now focus on \textit{net zero flux} ($\Phi=-1$), from which the zero-flux Chern bands emerge.
We plot the band structure (red) in Fig.~\ref{fig:figthree}b, together with
the adiabatic approximation bands (blue). \add{The adiabatic approximation becomes less accurate for increasing twist angle \cite{macdonaldMorales-DuranMagicAnglesFractional2024}. However, it still captures essential features of the full band structure at $\theta=1.67^\circ$.
For increasing twist angle, the electrons can escape the potential traps formed by the 
adiabatic potential $\tilde V(\mathbf r)$
and the kinetic energy term in Eq.~\eqref{eq:sphammacdo} becomes the dominant energy scale. As a result, the adiabatic band structure becomes 
describable in the nearly-free electron (weak periodic potential) approach. 
Indeed, the adiabatic bands (blue) are close to the backfolded bare
kinetic energy $-\frac{\hbar^2 \mathbf k^2}{2m^*}$ (dashed green in Fig.~\ref{fig:figthree}b), with degeneracies lifted by the periodic potential and the remnant effective magnetic field.} 

Like in the tight-binding regime, the Chern bands at $\Phi=0$ emerge as Landau levels from the parabolic band maximum at
$\Phi=-1$. 
However, as understood above, at $\Phi=-1$, the band structure in fact resembles backfolded free-electron bands, 
valid up to a filling of $\nu=1$ electron per unit cell. This provides enough states for the zeroth Landau level to persist to zero flux, $\Phi=0$, and create a band suitable for fractional quantum Hall liquids. For the second Chern band, on the other hand, the deviations of the $\Phi=-1$ band structure from pure parabolic
lead to a notable dispersion at $\Phi=0$.

To connect the present discussion to that of Section~\ref{sec:tightbindingregime}, we use the nearly-free electron picture to confirm that the Haldane-like topology of the topmost
two bands at flux $\Phi=-1$ remains stable up to large twist angles. In App.~\ref{app:nearlyfreeelectrons}, 
we obtain the topology of the topmost bands using symmetry indicators \cite{bernevigFangBulkTopologicalInvariants2012} in the adiabatic picture.
We find that the Haldane sequence of Chern numbers $(+1,-1)$ is remarkably robust, persisting up to 
$\theta\approx 1.96^\circ$. At this twist angle, there is a band crossing of the second and third bands at the $K_M$ and $K_{M'}$ points, which changes the Chern number of the second band to $C=-3$. 
We find that in the full continuum model, this topological phase transition occurs at $\theta \approx 2.02^\circ$, 
in surprisingly good agreement given the worsening of the adiabatic approximation for increasing twist angles.
We show the bands in the $(+1,-3)$ regime in Fig.~\ref{fig:figthree}c for a twist angle of $\theta=2.2^\circ$, exhibiting strong deviations between the full (red) and the adiabatic bands (blue).

\begin{figure*}[t]
    \includegraphics[width=1\textwidth]{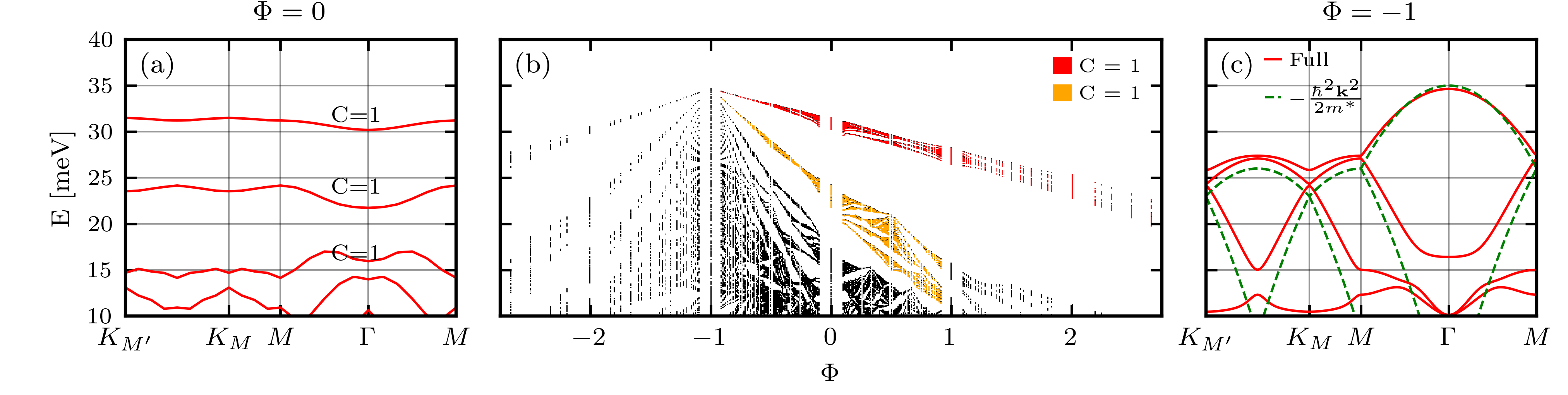}
    \caption{(a) Zero field band structure for a realistic model of twisted bilayer \mote{} 
    at twist angle $\theta =2.1^\circ$. We include second harmonic terms following \cite{choAhnFirstLandauLevel2024,xiaoZhangPolarizationdrivenBandTopology2024}.
(b) Energy spectra as a function of  magnetic field for the same system, highlighting the topmost $C=1$ band in red and second topmost $C=1$ band in orange.
(c) Band structure at $\Phi=-1$ (red) and backfolded free parabolic bands of the underlying \mote{} (dotted green), shifted by a constant in energy. 
\add{Note that the topmost branch of the backfolded free electron bands are doubly degenerate along the $K_{M'}-K_M-M$ line in the Brillouin zone, cf. the discussion in Appendix
~\ref{app:parabolicdegeneracies}.}
 }\label{fig:figfour}
\end{figure*}

\section{Twisted \mote{} at $\theta=2.1^\circ$: emergent nearly free electrons at flux $\Phi=-1$.}
\label{sec:mote}

Very recently, twisted bilayer \mote{} has been argued to host first Landau level analogs with non-Abelian states
\cite{makKangEvidenceFractionalQuantum2024,choAhnFirstLandauLevel2024,xiaoWangHigherLandauLevelAnalogues2024,fuReddyNonAbelianFractionalizationTopological2024,zhangXuMultipleChernBands2024,xiaoZhangPolarizationdrivenBandTopology2024}. At $\theta = 2.1^\circ$, density functional theory \cite{choAhnFirstLandauLevel2024,xiaoZhangPolarizationdrivenBandTopology2024}
predicts the three topmost bands to have the same Chern number, with the two top bands well detached from the remainder of the spectrum, see Fig.~\ref{fig:figfour}a for a continuum band structure. 
In order to fit the ab initio band structure, second-harmonic terms need to be included \cite{choAhnFirstLandauLevel2024,xiaoZhangPolarizationdrivenBandTopology2024,zhangXuMultipleChernBands2024}, see Sec.~\ref{sec:bandstructuresintro} for details. The energy level diagram as a function of magnetic field, shown in Fig.~\ref{fig:figfour}b, is qualitatively similar to that of \wse{} at larger twist angles (cf.\ Fig.~\ref{fig:figthree}a), with the two topmost Chern bands (highlighted in red and orange) clearly emerging from flux $\Phi=-1$. Importantly, the second Chern band (shown in orange) is clearly detached for \mote{}. 

As for \wse{}, we can obtain insight into the origin of these topological bands by studying the $\Phi=-1$ band structure. As  
shown in red in Fig.~\ref{fig:figfour}c, the band structure exhibits a distinct non-degenerate parabolic band maximum. Remarkably, a comparison to the bare TMD dispersion $-\frac{\hbar^2 \mathbf k^2}{2m^*}$ (dashed green lines) reveals a striking similarity with the full band structure for the two topmost bands.
\add{Note that the double degeneracy of the topmost branch of the free-electron bands along the $K_{M'}-K_M-M$ path in the Brillouin zone is lifted in the full band structure.}

Since the bands at flux $\Phi=-1$ are almost parabolic up to a filling of $\nu=2$, we expect at least two Landau-level-like bands when adding one flux quantum per unit cell. This is in agreement with recent arguments for the band structure at zero applied flux \cite{choAhnFirstLandauLevel2024}. We note that due to the rapidly varying potential profile, the adiabatic approximation breaks down for this model.
Nevertheless, the bands at \textit{net zero flux} ($\Phi=-1$) remain describable within the nearly free electron model similar to Eq.~\eqref{eq:sphammacdo}. It is this feature of the $\Phi=-1$ band structure that implies that the two topmost bands at zero flux ($\Phi=0$) can be understood as the two lowest Landau levels of a nearly free electron system. As a consequence, many remarkable phenomena associated with the fractional quantum Hall effect are likely to carry over.

The most important difference from the fractional quantum Hall effect is the 
presence of a time-reversed partner in the $K'$-valley.
This allows the possibility of time-reversal-symmetry preserving 
topological order, which has recently been proposed \cite{makKangEvidenceFractionalQuantum2024}.

We comment on Ref.~\cite{fuReddyNonAbelianFractionalizationTopological2024}, 
which used the adiabatic approximation bands for a model without higher harmonics to argue for first-Landau level physics.
As we have seen, the adiabatic approximation for larger twist angles naturally predicts Landau levels, as the $\Phi=-1$ band
structure becomes nearly free. However, the approximation itself breaks down for these large twist angles.  We have shown however, without 
invoking the adiabatic approximation, that a realistic model of \mote{} leads to nearly free bands at $\Phi=-1$. 

\section{Discussion}
In summary, we studied twisted transition metal dichalcogenide bilayers in an out-of-plane magnetic field. 
We first studied a minimal model of twisted \wse{} in the localized, Haldane regime, in which the magnetic field spectra qualitatively
match those of the Haldane model in a magnetic field. We showed that compared to the zero flux case, 
the bands at the \textit{point of net zero flux} ($\Phi=-1$) admit a Haldane description for an extended range of twist angles. The zero flux $C=1$ bands emerge as Landau levels from the bands at negative unit flux quantum per unit cell ($\Phi=-1$). At larger twist angles, the bands at $\Phi=-1$ resemble electrons in a weak periodic potential, a fact naturally explained in the adiabatic approximation.

For a realistic model of \mote{}, we found that the bands at $\Phi=-1$ are essentially obtained by backfolding a parabolic dispersion,
giving a clear physical picture for the second topmost band at zero flux as the first Landau level of this dispersion. 
This finding is of immediate relevance for the study of fractionalized states, as it provides a simple picture for the zeroth and first Landau level nature of the bands at $\Phi=0$.
This allows the phenomenology from the fractional quantum Hall effect to be translated
into the present context, with the crucial enrichment that in TMDs, there are two time-reversed copies, rendering the fractional quantum spin Hall effect a possibility 
\cite{makKangEvidenceFractionalQuantum2024}. 

For half filled TMD bands, experiment suggests an interpretation in terms of a composite Fermi liquid \cite{xuParkObservationFractionallyQuantized2023}. In fact, to lowest order, the mean-field band structure of composite fermions at half filling is precisely the $\Phi=-1$ band structure. Thus, our results imply that they have a strongly dispersive nature, which is a necessary condition for a composite fermion liquid to occur \cite{cooperMollerFractionalChernInsulators2015,parkerDongCompositeFermiLiquid2023,fuGoldmanZeroFieldCompositeFermi2023a,fuSternTransportPropertiesHalffilled2023,santosLuFractionalChernInsulators2024}. 

We now comment on the experimental relevance of our findings. In actual samples both, the $K$ and $K'$ valleys coexist and doped holes 
distribute among them by optimizing the single particle and interaction energies. At finite applied flux, the single particle spectra of the two valleys differ drastically -- the energy levels in the $K'$-valley at flux $\Phi$ are the same as the energy levels in the $K$-valley at flux $-\Phi$. While immaterial for and hence not included in our single-valley calculations, Zeeman splitting is significant when considering both valleys of TMDs \cite{marieRobertMeasurementConductionValence2021}, giving an additional energy splitting between the valleys at finite magnetic field. Thus, the precise way holes distribute among the two valleys is a subtle interplay of orbital, Zeeman, and interaction effects \cite{vafekWangInteractingPhaseDiagram2024}.
In the small twist angle regime with valley polarization, the Haldane-like Hofstadter spectrum should be observable in magnetotransport, with Landau fans emerging from integer applied fluxes. A tantalizing possibility would be the observation of the topological phase transition between fluxes $\Phi=-1$ and $\Phi=-2$ per unit cell, at which two Landau fans collide.
For larger twist angle \wse{} and \mote{} at $\theta=2.1^\circ$, we recover a parabolic dispersion in the $K$-valley at flux $\Phi=-1$ coexisting with a flat detached $C=1$ band in valley $K'$. The reentrant free electron  band of valley $K$ should be observable when the chemical potential is inside a gap for valley $K'$.

Our study paves the way to analyze arbitrary Chern bands and establish their 
connection to Landau levels by studying the Hofstadter spectra.
For $C=\pm 1$ bands, one has to trace their evolution between $\Phi=-1$ and $\Phi=0$.
Provided the band of interest can be traced all the way to $\Phi=-1$, 
the way it emerges from the band structure at $\Phi=-1$ allows it to be identified as a Landau level. The ideality of this level at $\Phi=0$ can be obtained from the similarity of the $\Phi=-1$ band structure with an elementary band structure, such as a Dirac cone or parabolic free electron band.
This provides a powerful complement to analytical approaches \cite{yangWangExactLandauLevel2021,vishwanathLedwithFractionalChernInsulator2020,ledwithFujimotoHigherVortexabilityZero2024,wangLiuTheoryGeneralizedLandau2024}.

As an interesting application, let us compare TMD band structures with twisted bilayer graphene in the chiral limit. For a given spin and valley flavor, there are two sublattice polarized zero-energy bands with opposite Chern numbers, a $C=1$ band on the $A$-sublattice and a $C=-1$ band on the $B$-sublattice. Thus, at $\Phi=1$, the $A$-sublattice band gains one state per unit cell,
while the $B$-sublattice band disappears.
Particle-hole symmetry necessitates this vanishing of the 
$C=-1$ band to be accompanied by a gap closing with the remote bands \cite{milekhinPopovHiddenWaveFunction2021,sternShefferChiralMagicangleTwisted2021,moraDattaHelicalTrilayerGraphene2024}. This enforces a Dirac cone in the band structure at $\Phi=1$. For fluxes $\Phi<1$, the $B$-sublattice band emerges as the zeroth Landau level of this Dirac cone. For $\Phi>1$, there is a Chern number $C=1$, $A$-sublattice polarized zeroth Landau level, so that the zero-energy 
bands have a total  Chern number of $C=2$ with $\nu = 2\Phi$ states per unit cell,
similar to the Landau level spectrum in the absence of a moir\'e tunneling term. 
The particle-hole symmetry and the ensuing emergence of the low-lying Chern bands from a Dirac cone explain why no analogs of higher Landau levels for a quadratic dispersion are present in twisted bilayer graphene. In contrast, for twisted TMDs, particle-hole symmetry is absent, allowing for a parabolic band extremum and enabling higher Landau levels to emanate from it.

The data sets used to generate the figures of this article are available from Zenodo \cite{moraKolavrDOI1052812024}. 

\begin{acknowledgments}
We thank E.\ Bergholtz, L.\ Glazman, D.\ Guerci, 
T.\ Holder, and A.\ Stern for helpful discussion. We gratefully acknowledge 
support of the research in Berlin by Deutsche Forschungsgemeinschaft through CRC 183 (project C02) and in Berlin as well as Paris by a joint ANR-DFG project (TWISTGRAPH, ANR-21-CE47-0018).  \end{acknowledgments}

\appendix
\section{Formulas for the Landau level basis}
\label{app:landauformulas}

The first term in the product in Eq.~\eqref{eq:matrixelementsep}
is a standard expectation value in Landau levels, and is given in terms of Laguerre polynomials 
 \cite{shankarMurthyHamiltonianTheoriesFractional2003}.
 For $n' \leq n$:
\begin{multline}
\label{eq:laguerreev}
\braket{n'|\exp\left(i \frac{\Pi \times \mathbf g}{B}\right)|n } =\\
(z^*)^{n'-n} \sqrt{\frac{n'!}{n!}}L^{n-n'}_{n'}(|z|^2)   \exp(-|z|^2/2),
\end{multline}
with $z = [(\mathbf g)_x + i (\mathbf g)_y] \frac{l_B}{\sqrt{2}}$ and $L^{n-n'}_{n'}(|z|^2)$ is the associated Laguerre polynomial.
For $n' > n$, we use 
$\braket{n'|\exp\left(i \frac{\Pi \times \mathbf g}{B}\right)|n}=\left[\braket{n|\exp\left(-i \frac{\Pi \times \mathbf g}{B}\right)|n' }\right]^*$.

For t;he second term in Eq.~\eqref{eq:matrixelementsep}, 
we write $\mathbf g = \localpha \mathbf G_1 + \locbeta \mathbf G_2$ to
separate the exponential
\begin{equation}
\exp\left(i \mathbf g \cdot \mathbf R\right) = 
c_1 \exp\left(i \localpha \mathbf G_1\cdot \mathbf R\right) \exp\left(i \locbeta \mathbf G_2 \cdot \mathbf R\right),
\end{equation}
where $c_1 =\exp (- \pi i \localpha \locbeta \frac{q}{p}) $ is the commutator from the Baker–Campbell–Hausdorff formula,
which states that $ e^{X+Y}=e^X e^Y e^{-\frac{1}{2}[X,Y]}$ for operators $X$ and $Y$ whose commutator $[X,Y]$ is a number.
At this point our basis definition, Eq.~\eqref{eq:llbasisdef} facilitates the evaluation.
For one, $\exp\left(i \locbeta \mathbf G_2 \cdot \mathbf R\right)$ changes $\mathbf k \to \mathbf k + \locbeta \mathbf G_2$.
If this momentum lies outside the Landau level Brillouin zone, compared to the basis definition of Eq.~\eqref{eq:llbasisdef}, there is an extra factor  
$\exp\left(i \lfloor k_2+ \locbeta/p \rfloor p \mathbf G_2 \cdot \mathbf R\right)$, where $\lfloor x \rfloor$ denotes the integer part of number $x$.
This operator is in fact proportional to a translation by $-q \mathbf a_1$ (using the relation of Eq.~\eqref{eq:relationmagnetictranseiqR}), and can be easily evaluated using the
translation properties of our basis states, Eq.~\eqref{eq:deflatticetranslations1}.
We have
\begin{eqnarray}
e^{i \locbeta \mathbf G_2 \cdot \mathbf R}\ket{\mathbf k}& =&e^{i \lfloor k_2+ \locbeta/p \rfloor p \mathbf G_2 \cdot \mathbf R} \ket{\mathbf k+ \locbeta \mathbf G_2}\nonumber\\
&=&(T_{-q \mathbf a_1})^{\lfloor k_2+ \locbeta/p \rfloor} \ket{\mathbf k+ \locbeta \mathbf G_2}\nonumber\\
&=&e^{-i 2\pi k_1 \lfloor k_2+ \locbeta/p \rfloor}  \ket{\mathbf k+ \locbeta \mathbf G_2}.
\end{eqnarray}
Next, we apply  
$\exp\left(i \localpha \mathbf G_1\cdot \mathbf R\right)$ 
leading to 
\begin{eqnarray}
e^{i \localpha \mathbf G_1 \cdot \mathbf R}\ket{\mathbf k+ \locbeta \mathbf G_2} 
= T_{\frac{q}{p}\localpha \mathbf a_2} \ket{\mathbf k+ \locbeta \mathbf G_2}\nonumber\\
= e^{i 2\pi  (k_2 p +\locbeta) \frac{q}{p}\localpha} \ket{\mathbf k+ \locbeta \mathbf G_2},
\end{eqnarray}
where we used Eq.~\eqref{eq:relationmagnetictranseiqR} in the first line.
Putting everything together:
\begin{multline}
 \braket{\mathbf k'|\exp\left(i \mathbf g \cdot \mathbf R\right)|\mathbf k} = \delta_{\mathbf k', \mathbf k + \locbeta \mathbf G_2} 
\exp (- i\pi  \localpha \locbeta \frac{q}{p})\\
\exp(-i 2\pi k_1 \lfloor k_2+ \locbeta/p \rfloor ) 
\exp\left\{i 2\pi  (k_2 p +\locbeta) \frac{q}{p}\localpha\right\},
\end{multline}
so that momentum is conserved up to multiples of $\mathbf G_2$.
The orbit consists of $p$ momenta $\mathbf k, \mathbf k+ \mathbf G_2, \ldots \mathbf k +(p-1)\mathbf G_2$, which need to be kept in the calculation.

In total, therefore, at flux $p/q$ per unit cell, at a given momentum
keeping $N_{LL}$ Landau levels,
we obtain a matrix of dimension $2 \cdot N_{LL} \cdot p$,
where $2$ is the number of layers.

\section{Degeneracies of backfolded parabolic bands}
\label{app:parabolicdegeneracies}
\begin{figure}[t]
    \centering
    \includegraphics[width=1\columnwidth]{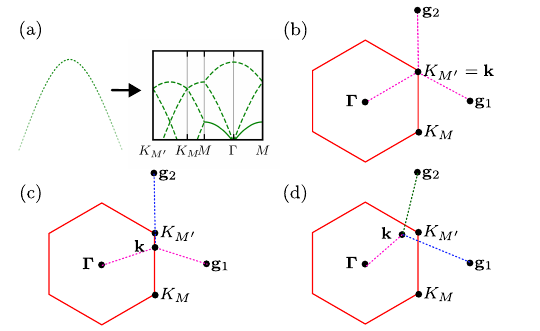}
    \caption{(a) Illustration of the backfolding of a free electron dispersion into the Brillouin zone.
    (b) For a point $\mathbf k =K_{M'}$ at the moir\'e Brillouin zone corner, there is a threefold degeneracy of the topmost branch of the 
    backfolded bands, as the distances of $\mathbf k$ to $\Gamma$, $\mathbf g_1$ and $\mathbf g_2$, shown in purple, are all equal.
    (c) For a point $\mathbf k$ along the $K_M-K_{M'}$ line, the degeneracy is partially lifted.
    The distances to $\Gamma$ and $\mathbf g_1$ (purple) are equal, while the distance to $\mathbf g_2$ (blue) is different.
    (d) For a generic point $\mathbf k$ in the Brillouin zone, there is no degeneracy, as
    the distances are all different.}
    \label{fig:parabolicdistances}
\end{figure}

\add{
The degeneracies of backfolded parabolic bands (Fig.~\ref{fig:parabolicdistances}a) in the moir\'e Brillouin zone
are best understood by examining copies of the parabolic dispersion shifted by reciprocal lattice vectors $\mathbf g$.
The value of that copy at a given point $\mathbf k$ in the Brillouin zone is proportional to the square of the
distance from the shift vector $\mathbf g$.
For example, around the $K_{M'}$ point, the three points $\Gamma$, $\mathbf g_1$, and $\mathbf g_2$, shown in Fig.~\ref{fig:parabolicdistances}b, 
are relevant for the topmost backfolded bands.
There is a threefold degeneracy at $\mathbf k = K_{M'}$. The distance of $K_{M'}$ from $\Gamma$, $\mathbf g_1$ and $\mathbf g_2$ is equal, being related by $C_{3}$ rotation around $K_{M'}$.
As pictured in Fig.~\ref{fig:parabolicdistances}c, for a point $\mathbf k$ along the $K_{M'}-K_M$ high-symmetry line, the distances to $\Gamma$ and $\Gamma+\mathbf g_1$ remain equal and decrease, while the distance to $\Gamma + \mathbf g_2$ increases. As a result, the threefold degeneracy at the  $K_{M'}$ point is lifted into a 2+1 degeneracy from top to bottom, see Fig.~\ref{fig:parabolicdistances}a, along $K_{M'}-K_M$. The same lifting occurs along $K_{M}-M$.
At a generic point $\mathbf k$ away from high symmetry lines (Fig.~\ref{fig:parabolicdistances}d), there is no degeneracy -- the distances from 
$\Gamma$, $\mathbf g_1$, and $\mathbf g_2$ are all different.}

\section{Details on the adiabatic model}
\label{app:adiabaticdetails}

For small twist angles, an adiabatic approximation for twisted TMDs was proposed in Ref.~\cite{yaoZhaiTheoryTunableFlux2020}.
To derive this picture, the full TMD model of Eq.~\eqref{eq:sphamtmd} is first
rewritten in the convention where momentum in each layer is measured with respect to its band maximum: 
\begin{equation}
\label{eq:sphammanewgauge}
H^{K}= -\frac{(\hbar \mathbf k)^2}{2m^*}\sigma_0 + \boldsymbol{\Delta}(\mathbf r)\cdot \boldsymbol{\sigma} +\Delta_0(\mathbf r)\sigma_0,
\end{equation}
with the Pauli $\sigma_{0,x,y,z}$ matrices acting in layer space.
In terms of the layer potentials and interlayer tunneling from Eq.~\eqref{eq:sphamtmd},
we have
\begin{eqnarray}
\Delta_0 =\frac{1}{2}[V_t(\mathbf r)+V_b(\mathbf r)]\\
\Delta_x =\mathrm{Re}\left[e^{i(\mathbf K^t-\mathbf K^b)\cdot \mathbf r} T(\mathbf r)\right] \\
\Delta_y =-\mathrm{Im}\left[e^{i(\mathbf K^t-\mathbf K^b)\cdot \mathbf r} T(\mathbf r)\right]\\
\Delta_z =\frac{1}{2}[V_b(\mathbf r)-V_t(\mathbf r)],
\end{eqnarray}
where the factor $e^{i(\mathbf K^t-\mathbf K^b)\cdot \mathbf r}$ is due to measuring momenta in
each layer with respect to their respective band maximum.

The effective Zeeman field in this model is given by $\boldsymbol{\Delta}(\mathbf r)$ with 
direction $\hat{\boldsymbol{n}}(\mathbf r) =\frac{\boldsymbol{\Delta}(\mathbf r)}{|\boldsymbol{\Delta}(\mathbf r)|}$,
and indeed the model is closely analogous to a model of electrons in a spatially varying spin texture \cite{taillefumierBrunoTopologicalHallEffect2004}.
Projecting onto the spinor tracking this Zeeman field at every point gives the adiabatic model:
\begin{equation}
\label{eq:sphammacdoapp}
H^{K}_{\text{Adiabatic}}= -\frac{(\hbar \mathbf k-e\tilde{\mathbf{A}}(\mathbf r))^2}{2m^*} + \tilde V(\mathbf r),
\end{equation}
where $\tilde{\mathbf{A}}(\mathbf r)$ is an emergent vector potential corresponding to an emergent inhomogeneous periodic magnetic field $\beff$ given 
in Eq.~\eqref{eq:beffdefmain}
and $\tilde V(\mathbf r)$ is an effective potential, which has both terms due to the effective Zeeman field and
due to the quantum metric of the adiabatic transformation:
\begin{equation}
\label{eq:veffdef}
\tilde V(\mathbf r) = - D(\mathbf r) + |\boldsymbol \Delta (\mathbf r)| + \Delta_0(\mathbf r),
\end{equation}
with $D(\mathbf r) =\frac{\hbar^2}{8m^*}\sum_{i=x,y}[\partial_{i} \mathbf{\hat n}(\mathbf{r})]^2 $. 
In this adiabatic approximation, the 3D inversion symmetry translates into a $C_{2z}$ symmetry,
which together with $C_{3z}$ combine to a $C_{6z}$ symmetry. $C_{2y}\timereversal$ is inherited from the original model. 

\section{Nearly free electrons at $\Phi=-1$}
\label{app:nearlyfreeelectrons}
We now turn to the nearly free electron approach for the adiabatic model of Eq.~\eqref{eq:sphammacdo} in 
an externally applied flux $\Phi=-1$,
where the dominant term is the kinetic energy, and the potential $\tilde V(\mathbf r)$ and vector potential $\tilde{\mathbf{A}}(\mathbf r)$ act as perturbations. 

\subsection{Definition of natural units and Fourier expansion}
We define a natural energy unit for terms arising from the kinetic term, evaluating for \wse{}:
\begin{equation}
\rydberg = \frac{\hbar^2 \km^2}{2m^*} = (\theta[^\circ]) ^2 \SI{4.3}{meV},
\end{equation}
where $\km =8\pi \sin(\theta/2)/(3a_0)$ is the magnitude of the moir\'e $K_M$ point in the brillouin zone,
which we use to define a dimensionless derivative $\tilde \partial_i=\frac{1}{\km} \partial_i $, such that $\km \tilde \partial_i = \partial_i$,
with the advantage that dimensionless derivatives of $\hat{\boldsymbol n}$ are twist angle independent. 
The effective potential is given in Eq.~\eqref{eq:veffdef}, and we separate it into two parts.
First, the kinetic potential term $D(\mathbf r)$ is written as:
\begin{eqnarray}
D(\mathbf r) = \rydberg \frac{1}{4} \sum_{i= x,y} (\tilde \partial_i \mathbf{\hat n})^2 \nonumber\\
=\rydberg \sum_{\mathbf g}   \delta_{\mathbf g} e^{i \mathbf g \cdot \mathbf r},
\end{eqnarray}
where $\delta_{\mathbf g}$ are twist-angle independent numbers depending only on model parameters.
Second, the Zeeman potentials can also be expanded:
\begin{eqnarray}
|\boldsymbol \Delta (\mathbf r)| + \Delta_0(\mathbf r)
=\sum_{\mathbf g}   \Delta_{\mathbf g} e^{i \mathbf g \cdot \mathbf r},
\end{eqnarray}
where $\Delta_{\mathbf g}$ has no twist angle dependence.
In total, the potential term can be written as a Fourier series as:
\begin{equation}
    \tilde V(\mathbf r) = \sum_{\mathbf g}   (-\rydberg \delta_{\mathbf g}  + \Delta_{\mathbf g}) e^{i \mathbf g \cdot \mathbf r}
    \end{equation}
Note that in this way, the kinetic potential term $D(\mathbf r)$ has the twist angle dependence explicitly
pulled out in the form of a factor of $\rydberg$.
The effective magnetic field is written  as
\begin{equation}
\label{eq:beffdef}
    \nabla \times \tilde{\mathbf{A}}(\mathbf r) =\beff = -\km^2\frac{\hbar}{e} \frac{1}{2} {\bm {\hat n}} \cdot \left(\tilde \partial_x {\bm {\hat n}} \times \tilde \partial_y {\bm {\hat n}} \right),
\end{equation}
where the minus sign compared to \cite{macdonaldMorales-DuranMagicAnglesFractional2024} is because in our convention $B>0$ gives $C=1$ for electrons.
Writing as a Fourier series, we obtain
\begin{equation}
\frac{1}{2} {\bm {\hat n}} \cdot \left(\tilde \partial_x {\bm {\hat n}} \times \tilde \partial_y {\bm {\hat n}}\right) = 
\sum_{\mathbf g}   \beta_{\mathbf g} e^{i \mathbf g \cdot \mathbf r}.
\end{equation}
At $\Phi=-1$ the effective magnetic field is on average cancelled by the external one, meaning we need to remove the $\mathbf g=0$ term in the sum
to obtain the total magnetic field experienced by the electrons:
\begin{equation}
\label{eq:beffdefminusone}
    B_{\text{tot}}(\mathbf r) = -\km^2\frac{\hbar}{e} \sum_{\mathbf g\neq 0 }   \beta_{\mathbf g} e^{i \mathbf g \cdot \mathbf r},
\end{equation}
where $\beta_{\mathbf g}$ is twist angle-independent.
This leads to the vector potential at $\Phi=-1$:
\begin{equation}
\tilde {\mathbf{A}}(\mathbf r) =-\km^2 \frac{\hbar}{e}\sum_{\mathbf g \neq 0 } \frac{i \mathbf g \times \hat{z}}{|\mathbf g|^2}  \beta_{\mathbf g} e^{i \mathbf g \cdot \mathbf r}.
\end{equation}
We now need to evaluate the vector potential terms in Eq.~\eqref{eq:sphammacdo}. 
As $|\beta_\mathbf{g}| \ll 1$, 
we neglect the diamagnetic term which contains factors of  $\beta^2$, obtaining
for the paramagnetic [linear in $\tilde {\mathbf A}(\mathbf r)$] matrix element
\begin{multline}
\label{eq:definitionparamagnetic}
\braket{\mathbf k_2|H_{\text{para}}|\mathbf k_1}=\frac{e \hbar}{2m^*} \braket{\mathbf k_2| \tilde{\mathbf{A}}(\mathbf r)\cdot \mathbf k +  \mathbf k \cdot \tilde{\mathbf{A}}(\mathbf r)|\mathbf k_1}\\
=-i\rydberg\frac{(\mathbf k_1+\mathbf k_2)\times (\mathbf k_2-\mathbf k_1)}{|\mathbf k_2-\mathbf k_1|^2} \beta_{\mathbf k_2-\mathbf k_1}\\
=-i\rydberg\frac{2 \mathbf k_1\times \mathbf k_2}{|\mathbf k_2-\mathbf k_1|^2} \beta_{\mathbf k_2-\mathbf k_1}.
\end{multline}

We use the symmetry indicators of band topology
\cite{bernevigFangBulkTopologicalInvariants2012}, which for a $C_{6z}$-symmetric system give the Chern number $C_i$ of a band $i$ as: 
\begin{equation}
\label{eq:chernnumberindicator}
   e^{i\pi C_i/3} = \eta_i(\Gamma)\theta_i(K_M)\zeta_i(M),
\end{equation}
where $\eta_i(\mathbf k)$, $\theta_i(\mathbf k)$ and $\zeta_i(\mathbf k)$ are
the $C_{6z}$, $C_{3z}$ and $C_{2z}$ eigenvalues of band $i$ at momentum $\mathbf k$.
In what follows, we evaluate the 
eigenvalues by considering the effective Hamiltonians inside the highest energy degenerate subspaces at high symmetry momenta. 
\subsection{Hamiltonian at the $M$-point}
At a given $M$ point there are only two relevant states, denoted $M^{(1)}$ and $M^{(2)}$ in Fig.~\ref{fig:kpointhoppings}a.
They are connected by a lowest magnitude reciprocal vector, giving a hopping
\begin{equation}
\label{eq:mpointhopping}
    t = \Delta_{1} -\delta_1E_0,
\end{equation}
which is necessarily real due to $C_{2z}$.
This gives energies 
\begin{equation}
\epsilon_{\pm} =  \pm (\Delta_{1} -\delta_1E_0),
\end{equation}
with $C_{2z}$ eigenvalues $\pm 1$. Above, we omitted a constant energy shift as it does not affect the ordering of the states.
For small twist angles (small $E_0$), the topmost band has $C_{2z}$ eigenvalue $-1$ (note that both $\Delta_1$ and $\delta_1$ are negative)
and the second band $+1$.
As twist angle increases, at the point
$\Delta_{1} -\delta_1E_0=0$, 
which happens at $\theta \approx 2.25^\circ$, the eigenvalues swap places.
Summarizing, for $\theta \lessapprox2.25^\circ$:
\begin{eqnarray}
\zeta_1(M) = -1\\
\zeta_2(M) = 1,
\end{eqnarray}
while for larger angles:
\begin{eqnarray}
\zeta_1(M) = 1\\
\zeta_2(M) = -1.
\end{eqnarray}

\subsection{Hamiltonian at the $K_M$-point: a three-site model}
\begin{table}
\begin{tabular}{c|c|c|c}
$i$ & $\Delta_i$ & $\delta_i$ & $\beta_i$ \\
\hline
1  & \SI{-2.61}{\meV}  & \SI{-0.12}{} & \SI{-0.11}{} \\
\hline
2  &  \SI{1.34}{\meV} & \SI{-0.06}{} &  \SI{-0.05}{}\\
\hline
3  & \SI{-1.13}{\meV}  &\SI{0.14}{}  & \SI{0.14}{} \\
\hline
\end{tabular}
\caption{Coefficients $\beta$,$\delta$ and $\Delta$ of the adiabatic model of \wse{}.}
\label{suptab:ckkprime}.
\end{table}

\begin{figure}[t]
    \centering
    \includegraphics[width=1\columnwidth]{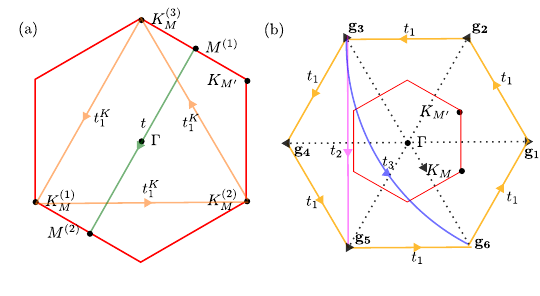}
    \caption{(a) Moir\'e Brillouin zone, showing two degenerate $M$ points $M^{(1)}$ and $M^{(2)}$ coupled by a hopping $t$
    and three degenerate $\km$ points, coupled by hopping $t^K_1$.
(b) Six relevant momenta for the second highest state at the $\Gamma$ point. They are coupled by 
nearest neighbor hopping $t_1$, next-nearest neighbor hopping $t_2$ and 
third nearest neighbor hopping $t_3$.}
    \label{fig:kpointhoppings}
\end{figure}

At the $K_M$ point, are only three relevant momenta, denoted $K^{(1)}$, $K^{(2)}$ and $K^{(3)}$ in Fig.~\ref{fig:kpointhoppings}a.
They are connected by a hopping $t_1^K$, and
the Hamiltonian is
\begin{equation}\label{triplet}
     H_{K_M} =  -E_0 +       
     \begin{pmatrix}
         0 & (t^K_1)^* &  t^K_1 \\ t^K_1  &0 & (t^K_1)^* \\
         (t^K_1)^* & t^K_1 &  0
    \end{pmatrix},
\end{equation}
where $t^K_1$ is given as 
\begin{equation}
\label{eq:kmhopping}
    t^K_1 = \Delta_{1} +E_0(-\delta_1 -i \frac{1}{\sqrt{3}}\beta_1).
\end{equation}
The eigenstates are labeled by their $C_{3z}$ eigenvalues, $1,\omega=e^{2\pi i/3},\omega^*=e^{4\pi i/3}$, and have energies (omitting an overall constant):
\begin{eqnarray}
\epsilon_{1} = 2 \left[\Delta_{1} -\delta_1 E_0\right] \\
\epsilon_{\omega} =-\Delta_{1} +E_0(\delta_1 +\beta_1) \\
\epsilon_{\omega^*} =-\Delta_{1} +E_0(\delta_1 -\beta_1) ,
\end{eqnarray}
that are equal to:
\begin{eqnarray}
\epsilon_{1}[\SI{}{meV}] = -5.22  +0.24 E_0 \\
\epsilon_{\omega}[\SI{}{meV}] =2.61 -0.23 E_0  \\
\epsilon_{\omega^*}[\SI{}{meV}] =2.61  -0.01E_0 .
\end{eqnarray}
Therefore, for small twist angle $\omega^*$ and $\omega$ states are the highest and second highest energy states, respectively.
The second highest state of eigenvalue $\omega$ crosses with the eigenvalue $1$ state at $\theta \approx 1.96^\circ$
Summarizing, for $\theta \lessapprox 1.96^\circ$:
\begin{eqnarray}
\theta_1(K_M) = e^{4\pi i/3}\\
\theta_2(K_M) = e^{2\pi i/3},
\end{eqnarray}
while for larger angles
\begin{eqnarray}
\theta_1(K_M) = e^{4\pi i/3}\\
\theta_2(K_M) = 1,
\end{eqnarray}

\subsection{Hamiltonian at the moir\'e $\Gamma$-point: a six-site model} The $\Gamma$ point extremum itself, which is part of the topmost band
 trivial representation of $C_{6z}$, giving $\eta_1=1$ for all twist angle.
For the second topmost state at the moir\'e $\Gamma$ point, six momenta are relevant, given
by $\mathbf g_1,\ldots , \mathbf g_6$, with $|\mathbf g_i| = \sqrt{3}\km$ for $i=1,\ldots 6$.
We plot them in Fig.~\ref{fig:kpointhoppings}b, together with the nearest, next-nearest and third-nearest neighbor hoppings $t_1$, $t_2$ and $t_3$.
The Hamiltonian at this point is:
\begin{equation}\label{hexat}
     H_{\km} =  -\sqrt{3} E_0 +       
     \begin{pmatrix}
         0 & t_1^* &  t_2^*&t_3&t_2&t_1\\ 
         t_1  &0 & t_1^*&t_2^*&t_3&t_2 \\
         t_2 & t_1 &  0&t_1^*&t_2^*&t_3\\
        t_3 & t_2 & t_1&0&t_1^*&t_2^*\\
         t_2^* &t_3 &  t_2&t_1&0&t_1^*\\
         t_1^* &t_2^* &  t_3&t_2&t_1&0
    \end{pmatrix},
\end{equation}
where $\khopping_1$ is given by 
\begin{equation}
\khopping_1 = \Delta_{1} +E_0(-\delta_1 -i \sqrt{3} \beta_1),
\end{equation}
with the only difference to $t^K_1$ in Eq.~\eqref{eq:kmhopping}
being the geometric factor $\sqrt{3}$ multiplying $\beta_1$. 
The other hoppings are:
\begin{eqnarray}
\khopping_2 = \Delta_{2} +E_0(-\delta_2 -i \frac{1}{\sqrt{3}} \beta_2)\\
\khopping_3 = \Delta_{3} +E_0(-\delta_3),
\end{eqnarray}
where we note that $\khopping_3$ is real due to $C_{2z}$ symmetry.

To solve, we use the $C_{6z}$ symmetry, so that solutions can be labelled by $C_{6z}$ eigenvalues
$e^{i\frac{\pi}{3}m}$, where $m$ is the angular momentum.
Firstly, $t_3$ causes a large splitting between different $C_{2z}$ sectors, 
caused by the dip of $\tilde V(\mathbf r)$ at the midpoint between AB and BA sites of the unit cell, 
and the magnitude of this splitting increases with twist angle.
Since $t_3$ is negative, the upper branch has $C_{2z}$ eigenvalue $-1$, corresponding to angular momenta $m=1,3,5$.
The energies of these states are (up to an overall constant):
\begin{eqnarray}
\epsilon_{m=1} =-t_3 +2 \mathrm{Re}\left[ (t_1-t_2^*)e^{i\pi/3} \right]  \\
\epsilon_{m=3} =-t_3 +2 \mathrm{Re}\left[-t_1 +t_2  \right]\\
\epsilon_{m=5} =-t_3 +2 \mathrm{Re}\left[ (t_1-t_2^*)e^{-i\pi/3} \right],
\end{eqnarray}
which we can numerically evaluate for \wse{} to be:
\begin{eqnarray}
\epsilon_{m=1} [\SI{}{meV}] \approx -t_3 + 2\cdot \left[ -1.97 -0.15 E_0\right]  \\
\epsilon_{m=3} [\SI{}{meV}] \approx -t_3 +2\cdot \left[3.95 -0.06 E_0 \right]\\
\epsilon_{m=5} [\SI{}{meV}] \approx -t_3 + 2\cdot \left[ -1.97 +0.2 E_0\right] ,
\end{eqnarray}
where we do not evaluate $t_3$ as it is the same for all states.
We see that the $m=3$ state is by far highest in energy up to $\theta=2.5^\circ$.
Therefore, we have for realistic angles:
\begin{eqnarray}
\eta_1(\Gamma) = 1\\
\eta_2(\Gamma) = -1,
\end{eqnarray}
\subsection{Topology}
Putting the results together, we have for
$\theta\lessapprox 1.96^\circ$ the Chern number sequence:
\begin{eqnarray}
C_1 =1 \mod 6 \\
C_2=-1 \mod 6.
\end{eqnarray}
For $\theta\gtrapprox 1.96^\circ$, we have
\begin{eqnarray}
C_1 =1 \mod 6\\
C_2=-3 \mod 6,
\end{eqnarray}
predicting the correct phase transition and obtaining the 
twist angle of the transition to reasonable accuracy compared with the full model, in which it happens at $\theta \approx 2.02^\circ$. From the symmetry-indicator analysis, we also infer that the gap closing  at $\theta \approx 2.02^\circ$ between the second and the third band occurs through two Dirac cones at $K_M$ and $K_{M'}$.

\IfFileExists{klibrary.bib}{\bibliography{klibrary.bib}}{\IfFileExists{../klibrary.bib}{\bibliography{../klibrary.bib}}{\bibliography{~/notes/3_moire/5_moire_ll/klibrary.bib}}}

\end{document}